\documentclass[pra,amsmath,superscriptaddress,a4paper,floatfix,twocolumn]{revtex4}  
\usepackage{amssymb,amsmath,verbatim,units,bbm,tikz,color}

\newcommand{\tr}{\mbox{tr}}
\newcommand{\ket}[1]{\left | #1 \right \rangle}
\newcommand{\bra}[1]{\left \langle #1 \right |}

\newcommand{\beq}{\begin{equation}}
\newcommand{\eeq}{\end{equation}}
\newcommand{\beqa}{\begin{eqnarray}}
\newcommand{\eeqa}{\end{eqnarray}}

\begin{document}
\title{Robust symmetry-protected metrology with the Haldane phase}
\author{Stephen D. Bartlett}
\affiliation{Centre for Engineered Quantum Systems, School of Physics, The University of Sydney, Sydney, NSW 2006, Australia}
\author{Gavin K. Brennen}
\affiliation{Centre for Engineered Quantum Systems, Macquarie University, Sydney, NSW 2109, Australia}
\author{Akimasa Miyake}
\affiliation{Center for Quantum Information and Control, Department of Physics and Astronomy, University of New Mexico, Albuquerque, NM 87131, USA}

\begin{abstract}
We propose a metrology scheme that is made robust to a wide range of noise processes by using the passive, error-preventing properties of symmetry-protected topological phases. The so-called fractionalized edge mode of an antiferromagnetic Heisenberg spin-1 chain in a rotationally- symmetric Haldane phase can be used to measure the direction of an unknown electric field, by exploiting the way in which the field direction reduces the symmetry of the chain.  Specifically, the direction of the field is registered in the holonomy under an adiabatic sensing protocol, and the degenerate fractionalized edge mode is protected through this process by the remaining reduced symmetry. We illustrate the scheme with respect to a potential realization by Rydberg dressed atoms. 
\end{abstract}
\date{25 October 2017}
\maketitle

\section{Introduction}

Precision measurements using engineered quantum systems are reaching new records in accuracy.  For example, electric field sensing using single quantum probes has been demonstrated using single electron transistors~\cite{Devoret}, spin qubits~\cite{Dial} and NV colour centres in diamond~\cite{Detal}, and magnetic field sensing using NV centres~\cite{Maze} and SQUIDs~\cite{Kirtley}.   The challenge with quantum metrology schemes based on single quantum systems is the considerable susceptibility to noise associated with such sensitive probes.  Recent work into quantum metrology schemes that make use of quantum error correction~\cite{AVAR2014,DSFK2014,KLSL2014,Ozeri2013,Zhou2017} attempts to address this issue.  The control requirements and experimental complexity for such schemes, however, are quite daunting with current experimental techniques.

There are quantum many-body systems that are naturally robust against certain errors.  In particular, symmetry protected (SP) topological phases support gapped ground spaces with a degeneracy that is protected against all noise that respects some symmetry group $G$~\cite{CGW2011}.  As such, systems prepared in SP phases can be used for hardware protected quantum information devices.  A holonomic processing for fully universal quantum computation using the properties of SP phases was proposed in Ref.~\cite{RMBB2013}; see also Refs.~\cite{BBMR2010,M2010,BFC2013,ESBD2012,EBD2012,WB2014,JM2015}. These phases are less robust than fully topologically ordered phases which are protected against \emph{any} local error.   On the other hand, such SP phases may be easier to engineer in the laboratory and in particular can exist in one-dimensional spin chains.  Indeed, the SP topological order of such phases also offers a unique perspective on information processing in such systems:  quantum information can be robustly encoded into such phases and is protected from all local errors that respect the symmetry, and in addition the symmetry group provides a `handle' by which to manipulate the quantum information using local interactions.

In this paper, we describe how SP phases can be used to perform metrology with an intrinsic robustness to a wide range of errors.  In comparison to schemes for metrology that use quantum error correction or other techniques with fast control~\cite{AVAR2014,DSFK2014,KLSL2014,Ozeri2013,Zhou2017,Sekatski2016}, our method removes the requirements for active error correction and the local control associated with it, instead using the passive error-preventing properties of the SP phase.  We present a sensing protocol that measures the direction of the field only, and then extend this protocol to measure both direction and strength.  The advantages of performing metrology in this setting are: \textit{(i) Immunity to timing errors.---}This measurement scheme uses holonomies so it is robust to timing errors in the gates.  In the protocol to measure the field direction only, the protocol is also immune to fluctuations in the strength of the field that is measured. \textit{(ii) Long storage times.---}The information gained from the field is stored in a gapped SP phase, which can hold the quantum information for a long time before decoherence sets in.  This protection could be advantageous in cases where measurement is very slow. \textit{(iii) Symmetry protection.}---The entire protocol respects the symmetry of the SP phase and is immune to error even during processing.  

\section{The Haldane phase and its symmetries}

Our construction makes use of the properties of SP phases in 1D spin chains.  An SP phase is defined to be a class of symmetric uniformly gapped Hamiltonians $H$ that are equivalent under symmetry-respecting adiabatic evolutions.  The class containing a symmetric product state is a trivial symmetric phase and other distinct classes are called symmetry-protected (SP) phases \cite{PTBO2010,CGW2011,SPC2011,PBTO2012,CGLW2012,HK2010,QZ2011}.  As a concrete example, we consider antiferromagnetically coupled spin-1 chains in the Haldane phase~\cite{haldane1983continuum}.  Consider such a chain with a bulk Hamiltonian with $SO(3)$ symmetry, i.e., that is rotationally invariant.  As a canonical representative of the bulk Hamiltonian, we use the nearest-neighbour Heisenberg coupling $H_{k,n} =J\sum_{j=k}^{n-1}\vec{S}_j\cdot \vec{S}_{j+1}$ for $J>0$, but we emphasise that our results are independent of the details of the Hamiltonian and apply equally to the entire Haldane phase. Several physical systems are available to engineer Haldane chains including polar molecules in optical lattices \cite{Polar:07}, trapped ions \cite{CR2014}, and trapped Rydberg dressed atoms \cite{Bijnen:15}.

For long enough chains, the Haldane phase is characterised by a four-dimensional degenerate ground space, with splitting exponentially small in the length. At the Heisenberg point, the correlation length is $\xi\approx 6.03$ and the gap to spin-$2$ excitations is $\Delta\approx 0.41 J$ \cite{Hagiwara:90,White:92}.   This degenerate ground space can be viewed as fractionalized spin-$1/2$ particles associated with each end of the chain.  For convenience, we focus our attention on just one of these edge modes, specifically the left edge mode. To remove the right edge mode from consideration, following Ref.~\cite{M2010}, for $n>\xi$, we can perform measurements on the right edge mode to decouple it.  The left edge mode qubit can be manipulated by a sequence of adiabatic coupling and decoupling of local Hamiltonian terms that are generic but respect particular symmetries~\cite{RMBB2013}, and which couple with the unknown field to be measured, in order to perform a field-dependent local adiabatic gate on the degenerate qubit state.   Tomographic techniques will then be used to infer information about the field based on the evolved state of this edge mode qubit.  

While our spin chain is $SO(3)$ invariant, the field we wish to measure will break this symmetry.  Consider a metrological scenario in which we wish to probe an electric field $\vec{E}_f=E_f\hat{m}_f$, where $E_f>0$ is the field strength, and $\hat{m}_f$ is a unit vector in the direction of the field.  The field direction $\hat{m}_f$ can define a reduced symmetry group, $D_\infty = SO(2) \rtimes Z_2 \subset SO(3)$, by combining this direction with an $SO(2)$ symmetry of rotations about it.  This is called the continuous dihedral group, which is a semi-direct product group consisting of arbitrary rotations about the axis $\hat{m}_f$ together with $\pi$-flips that invert this axis and thus is isomorphic to $O(2)$.  While this field breaks the full $SO(3)$ rotational symmetry of the bulk, the remaining $D_\infty$ symmetry preserves the properties of the Haldane phase including the degenerate edge modes.  We can enforce this $D_\infty$ symmetry in the sensing protocol by ensuring that the coupling of the field to the spin chain respects the symmetry.  For example, consider the coupling between the field and one of the spin-1 particles to be given by $H_f=J_f (S^{\hat{m}_f})^2$, where $J_f=J_f(E_f)>0$ describes the local spin coupling strength as a function of the electric field magnitude.  As such, the spins couple only to the axis defined by $\pm\hat{m}_f$.  The coupling of this form is a standard one describing coupling of a spin-1 particle to a vector field, preserving time reversal symmetry. This occurs, for example, (i) in the ground electronic states of NV centres in diamond interacting with an electric field~\cite{Detal}, (ii) the interaction of the atomic electric quadrupole moment of ions with an external electric quadrupole fields~\cite{Roos06}, (iii) alkali atoms interacting via the static or AC stark effect, and (iv) for systems interacting with an effective electric field like superconducting fluxonium qutrits encoded in phase eigenstates interacting with microwave fields~\cite{Brennen16}.  This interaction has the desired $D_\infty$ symmetry; note that only the symmetry of the interaction is important, and the details of the coupling Hamiltonian are irrelevant. 

We emphasise here that the symmetry reduction of this model, from the full $SO(3)$ rotation symmetry of our spin chain down to the $D_\infty$ symmetry for the coupling $H_f$, is determined by the direction of the unknown field $\vec{E}_f$.  That is, the field reduces the larger symmetry to a smaller one, and ultimately it will be the aim of our sensing protocol to determine the specific way in which this symmetry reduction occurs.  The Haldane phase is in an ordered SP phase under this reduced symmetry group $D_\infty$, which ensures the robustness of our sensing protocol.  Because the direction of the unknown field $\hat{m}_f$ is not known \emph{a priori}, we require a spin chain with the full $SO(3)$ rotation symmetry.  That is why hypothetically if the field were coupled to the spin-1 particle at the end of the chain through a coupling proportional to $\vec{E}_f \cdot \vec{S}$, then the system would possess only a $SO(2)$ symmetry.  However, this limited symmetry is not sufficient to protect the nontrivial SP phase of our spin chain.

(We note that rotationally-invariant states of spin chains have been proposed for metrology schemes in a different context~\cite{Toth2013}.  In this related work, a rotationally-invariant but topologically trivial state without gap protection allows for sensing of spatial gradients of fields and is insensitive to homogeneous fields.  In contrast, our metrology scheme makes use of rotationally-invariant, SP \emph{nontrivial} gapped ground states for homogenous field sensing.)

\section{Holonomic field sensing}

\begin{figure}
\begin{center}
	\includegraphics[width=\linewidth]{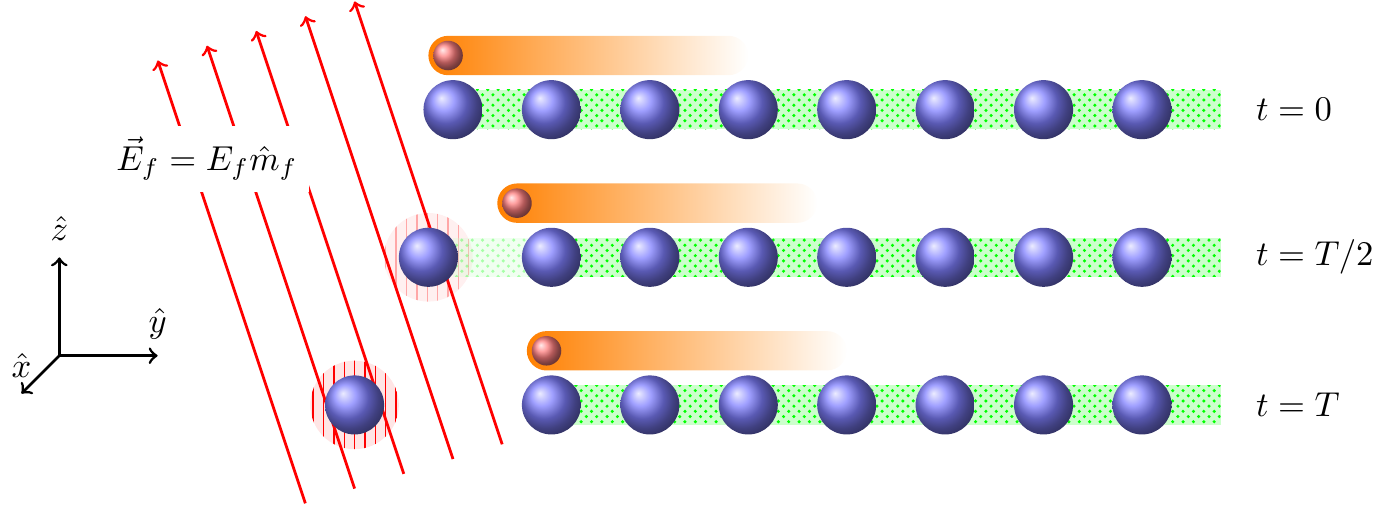}
\end{center}
	\caption{
	\label{fig:SensingAction}
	Basic action of the sensing operation.  Large (blue) spheres denote the spins of the chain.  The left edge carries a fractionalized edge degree of freedom (orange).  Adiabatically decoupling the boundary spin from its immediate neighbour while simultaneously subjecting it to interaction with the local field (red), transfers the encoded qubit to the slightly shorter chain while effecting a $\pi$ rotation about the local field axis.
	}
\end{figure}

\begin{figure}
\begin{center}
	\includegraphics[width=\linewidth]{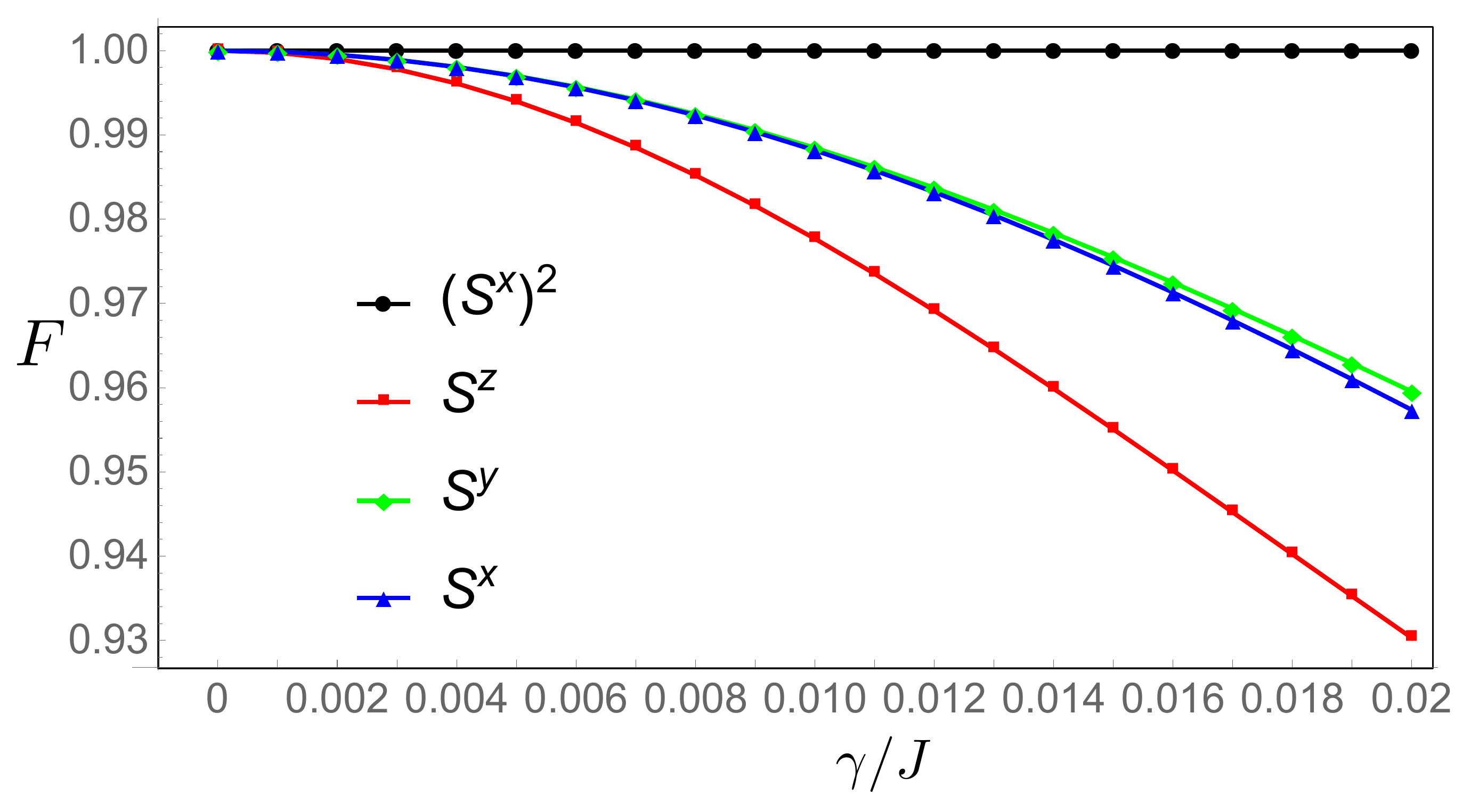}
\end{center}
	\caption{
	\label{fig:Fidelity}
Performance of the holonomic sensing protocol for measuring a field in the $\hat{m}_F=\hat{x}$ direction in the presence of perturbations to the Heisenberg coupled chain. Plotted is the fidelity $F=(\tr[\sqrt{\sqrt{\rho}\sigma\sqrt{\rho}}])^2$, to perform a $\pi$ rotation about the $\hat{x}$ axis, i.e. a bit flip, on the logical qubit encoded in the chain. 
Each point corresponds to the simulation of the holonomic gate for a chain of length $n=8$, over a time $T=10/J$, with the chain Hamiltonian $H(t)_{1,n}+H'$ where the perturbation is $H'=\gamma\sum_{j=1}^n O_j$, for the operator $O$ indicated.}
\end{figure}

We now detail the steps of the basic sensing operation in our scheme, as illustrated in Fig.~\ref{fig:SensingAction}.
We initiate the system in a ground state of the Hamiltonian $H_{k,n}$, starting with $k=1$. 
We then adiabatically turn off the interaction of spins $k$ and $k+1$, decoupling spin $k$ from the chain, while simultaneously allowing this spin to interact with the electric field through the interaction $H_f$.  We consider the parameterised family of Hamiltonians
\beq
  \label{eq:adiabaticshift}
  H_{k,n}(t)=f(t) J_f (S^{\hat{m}_f}_k)^2 +  g(t) J \vec{S}_k\cdot \vec{S}_{k+1}+H_{k+1,n} \,,
\eeq
with monotonic $f,g$ satisfying $f(0)=g(T)=0$ and $f(T)=g(0)=1$.  (Again, the precise Hamiltonian details are not relevant.)  This evolution decouples the left boundary spin from the bulk and prepares that spin in a unique ground state of $(S^{\hat{m}_f}_k)^2$.  As an example, the end spin could be dragged into the region of the field while the coupling with the chain is reduced.  The end spin is now disentangled from the slightly-shortened chain.  While the disentangled end spin contains information about $\hat{m}_f$, this information is not protected in any way.  Fortunately, the direction $\hat{m}_f$ is also encoded in the protected edge mode of the remaining chain, and it is this protected information that we will use in our protocol.  We note that the protocol is robust to the spatial extent of the field $\vec{E}_f$; this field can extend through the chain itself (coupling with spins as $J (S^{\hat{m}_f})^2_j$) as its symmetry preserves the Haldane phase, provided it is weak enough that it does not close the gap (see Fig. \ref{fig:Fidelity}).

The time-dependent Hamiltonian of Eq.~\eqref{eq:adiabaticshift} is invariant under the $D_\infty$ symmetry defined above.  As we now demonstrate using symmetry arguments, the resulting unitary action on the qubit degree of freedom in the edge mode, which is now squeezed onto a slightly shorter chain, is a $\pi$-rotation about the axis defined by $\hat{m}_f$.  To see this, we make use of two conserved quantities: 
\begin{align}
   i  \Sigma^{\hat{m}_f}_{k,n}&=\Big(\otimes_{j=k}^n \exp({i\pi S^{\hat{m}_f}_j})\Big)\otimes \exp\left(i\tfrac{\pi}{2}\sigma^{\hat{m}_f}\right) \,, \\
   i  \Sigma^{\hat{m}^\perp_f}_{k,n}&=\Big(\otimes_{j=k}^n \exp({i\pi S^{\hat{m}^\perp_f}_j})\Big)\otimes \exp\left(i\tfrac{\pi}{2}\sigma^{\hat{m}^\perp_f}\right) \,,
\end{align}
where $\hat{m}^\perp_f$ is any orthogonal vector to $\hat{m}_f$ (the choice of this orthogonal vector is arbitrary, due to $SO(2)$ invariance).  The rightmost operators $\sigma^{\hat{m}_f}$ and $\sigma^{\hat{m}^\perp_f}$ span a spin-$1/2$ subspace of the right boundary system, represented the other (unused) edge mode.  The operators $i\Sigma^{\hat{m}_f}_{k,n}$ and $i\Sigma^{\hat{m}^\perp_f}_{k,n}$ correspond to spin-1/2 representations of $\pi$-rotations about the vectors $\hat{m}_f$ and $\hat{m}_f^\perp$, and generate a $D_2 \subset D_\infty$ subgroup of transformations on the qubit degree of freedom associated with the left edge mode.  These operators commute with the family of Hamiltonians $H_{k,n}(t)$ as a result of the $D_\infty$ invariance.  We note that for a spin-1 representation we have $\exp(i\pi S^{\hat{m}^\perp_f})\ket{S^{\hat{m}_f}{=}0}=-\ket{S^{\hat{m}_f}{=}0}$, and this transformation property is the key to the basic sensing transformation.  Imagine the qubit initiates in a $+1$ eigenstate of $\Sigma_{k,n}^{\hat{m}_f}$, i.e., the state 
\begin{equation}
  |\psi(0)\rangle=|\mathbf{0}\rangle_{k,n}\equiv|\Sigma_{k,n}^{\hat{m}_f}{=}{+}1,H_{k,n}=0\rangle\,. 
\end{equation}  
After the adiabatic dynamics it becomes 
\begin{equation}
  |\psi(T)\rangle = |\Sigma_{k,n}^{\hat{m}_f}{=}{+}1,(S_k^{\hat{m}_f})^2{=}0,H_{k{+}1,n}{=}0\rangle\,.
\end{equation}  
But due to the product form of the set of operators $\Sigma_{k,n}^{\hat{m}_f}$ for $k=1,2,\ldots$, this is nothing other than the state $|S^{\hat{m}_f}{=}0\rangle\otimes |\Sigma_{k+1,n}^{\hat{m}_f}{=}{+}1,H_{k+1,n}{=}0\rangle$, meaning $|\psi(T)\rangle=|S^{\hat{m}_f}{=}0\rangle\otimes|\mathbf{0}\rangle_{k+1,n}$, up to some unknown phase. Following a similar argument, $|\mathbf{1}\rangle_{k,n}$ evolves to $|S^{\hat{m}_f}{=}0\rangle\otimes|\mathbf{1}\rangle_{k+1,n}$, up to a possibly different phase. 

To determine the relative phase accumulated by $|\mathbf{0}\rangle$ and $|\mathbf{1}\rangle$, we use the other conserved quantity $\Sigma^{\hat{m}^\perp_f}_{k,n}$.  Consider the time evolution of a logical qubit initialized in the $+1$ eigenstate of $\Sigma_{k,n}^{\hat{m}^\perp_f}$, which we denote $|\mathbf{+}\rangle_{k,n}$. Because $\exp(i\pi S^{\hat{m}^\perp_f})|S^{\hat{m}_f}{=}0\rangle=-|S^{\hat{m}_f}{=}0\rangle$, the eigenvalue of $\Sigma_{k+1,n}^{\hat{m}^\perp_f}$ in the final step must be $-1$, so $|\mathbf{+}\rangle_{k,n}$ is transformed into $|S^{\hat{m}_f}{=}0\rangle\otimes|\mathbf{-}\rangle_{k+1,n}$, where $|\mathbf{-}\rangle_{k+1,n}$ is the $-1$ eigenstate of $\Sigma^{\hat{m}^\perp_f}_{k+1,n}$. Therefore, the dynamics effects a $\pi$ rotation of the qubit about the $\hat{m}_f$ axis.  (We note that this argument is independent of the choice $\hat{m}^\perp_f$ orthogonal to $\hat{m}_f$.)

It is this robust unitary $\pi$-rotation about $\hat{m}_f$ on the protected edge mode qubit that will form the basic action of our metrology schemes.  We note that this basic action --- a rotation of a qubit by a fixed rotation angle of $\pi$ about an unknown axis --- is distinct from the basic action in a standard phase estimation scheme involving a rotation by an unknown angle about a fixed axis.

Initialization and readout of the qubit degree of freedom can be done following the techniques of Refs.~\cite{M2010,RMBB2013}.  Decoupling a boundary spin is performed as described above but without a local field, followed by a measurement of $S^{\hat{n}}$ for some direction $\hat{n}$, yields a nondeterministic measurement of the edge mode qubit in the $|{+}\hat{n}\rangle$, $|{-}\hat{n}\rangle$ basis corresponding to the outcome $m= \pm 1$.  (The outcome $m= 0$ does not measure the edge mode qubit, but rather rotates it by $\pi$ about the $\hat{n}$ axis.  If this outcome occurs, the measurement must be attempted again.)  Initialization into any desired qubit state can be performed using such a measurement, repeated until the desired $+1$ outcome is obtained.  We note that the gap closes during this process (the decoupled spin becoming degenerate) and so the gap protection is lost at the measurement step, as it must be.

\section{Metrology for the field direction}

Based on the holonomic field sensing described above, we now introduce a basic metrology scheme that measures the axis $\hat{m}_f$ but not the magnitude of the field $\vec{E}_f$.  Our basic action implements a $\pi$-rotation of our protected edge mode qubit about $\hat{m}_f$, which is then interrogated to acquire information about the direction $\hat{m}_f$.  We then repeat with a number $N$ of independently prepared edge modes. Each edge mode qubit is initialized in a fixed state $|\psi_0\rangle$, with Bloch vector $\hat{\psi}_0$. We then perform the above adiabatic step which rotates this state by $\pi$ about the unknown axis $\hat{m}_f$, and finally to perform tomographic measurements of the edge mode qubit to determine the final state. Note that, for a given initial Bloch vector $\hat{\psi}_0$, a tomographically-reconstructed final state $\hat{\psi}'$ uniquely determines the axis (not direction) of $\hat{m}_f$.

The precision of this estimate will depend on the specifics of the tomography scheme, as well as the (unknown) relationship between $\hat{\psi}_0$ and $\hat{m}_f$.  Because of the protected nature of the encoded qubits, the tomographic regime of interest is one of nearly-pure states uniformly distributed on the Bloch sphere, and so we look to the single-step adaptive protocol of Ref.~\cite{MRDFBKS13}.  This scheme can estimate $\hat{\psi}'$ to a precision $1-F = O(1/N)$ for $N$ samples, using a single adaptive step after around half the samples. This adaptive step can also be used, if needed, to change the initial state $\hat{\psi}_0$ to be at right angles with the estimated $\hat{m}_f$ at that stage, so as to have the maximum action of the $\pi$-rotation on this state. Because the fidelity measures state overlap, this implies a variance in the estimate of the field direction of $O(1/\sqrt{N})$.   

We quantify resources in this scheme using the number of independent samples $N$.  Each sample requires a fixed amount of time, determined by the requirement of adiabaticity, and so our resource measure could also be total time.   By this measure of resources, the scheme above performs at the standard quantum limit  (variance of $O(1/\sqrt{N})$) because it uses the $N$ protected edge mode spins independently.  Note, because of this independence, the scheme can be parallelized. Heisenberg-limited (variance of $O(1/N)$) schemes can also be developed using the basic sensing action defined here, through the use of entangled states or measurements on the edge modes of $N$ chains, e.g., GHZ states, and is also parallelizable. If we make the assumption that the unknown field is the same over all the $N$ boundary spins, then the resulting reduced $D_\infty$ symmetry is the same for all chains $N$ chains and protects a holonomic CPHASE gate~\cite{WB2014}.  With such a gate acting on nearest neighbour chains, a GHZ state of $N$ boundary spins can be prepared in constant depth.  As an additional remark, the boundary spins are eventually decoupled to a product state in this Heisenberg-limited scheme using the GHZ state. Thus, measuring disentangled end spins instead does not provide the same improvement of precision. 

Note that if the field strength $J_f$ is very weak, compared with say the temperature $kT$ or the energy scales of other error processes, then the gap on the boundary spin may not be sufficiently large to offer gap protection.  In addition, the adiabatic gate time can become very long.   In such a case, the basic sensing protocol above can be modified by the addition of a known strong field, used as reference, and treating the unknown field as a weak perturbation.   Consider a similar control schedule as before but where in addition to the field $\vec{E}_f=E_f\hat{m}_f$, there is a ``background" field $\vec{E}_b=E_b\hat{m}_b$ with strength $E_b>0$ and direction $\hat{m}_b$.  We will assume that $E_b\gg E_f$. This can be verified by other, less precise, measurements of the field, and will allow us to obtain both the axis and direction of $\hat{m}_f$ since it is then a perturbation on the known background field direction.  Additionally, this condition insures that the local gap of magnitude $J_{\rm tot}$ is large.  The total local interaction is
$H_{\rm local}=J_{\rm tot} (S_1^{\hat{m}_{\rm tot}})^2$.  

This technique additionally allows for the measurement of both the field \emph{strength and direction} of $\hat{m}_f$.  We will exploit the freedom to choose the direction of the background field during measurement runs.  For example, two rounds can be used to obtain estimates of the total field directions $\hat{m}_{\rm tot,1}$ and $\hat{m}_{\rm tot,2}$ for two known background fields $\hat{m}_{b,1}$ and $\hat{m}_{b,2}$.  (We assume that the unknown field is constant for the total duration of the two runs.)  With this information, the known strengths and directions of $\vec{E}_{b,1}$ and $\vec{E}_{b,2}$, and using the gate tomography scheme above, both the magnitude and direction of $\vec{E}_f$ can be determined.  Again, the scheme using $N$ independent edge modes will be standard quantum limited, but entangled schemes can increase this scaling.

\section{Performance}

We now consider how the performance of our scheme will compare with more traditional methods in quantum metrology.  We make the assumption that $J_f \sim J$, that is, that the field strength is comparable to the Heisenberg coupling strength such that neither strength is dominating the adiabatic gate sensing time.  With this assumption, we note that in the absence of noise, our scheme will perform comparably with a `traditional' approach based on using a single spin-1 particle with $H_f = J_f (S^{\hat{m}_f})^2$ as in, for example, Ref.~\cite{Detal}. This is because the time to perform the measurement is essentially the same for both approaches and the number of spins in the chain is finite ($n\sim 10$), so even if all the spins in the chain were instead used to probe the field, the precision is similar. 

Importantly however, in the presence of noise our robust scheme can outperform these other approaches.  It has been shown that for local noise, standard metrology using entangled states does essentially no better than using non-entangled states~\cite{Huelga,Escher:11,DD:12} (although see~\cite{Huelga2:2014,Chaves}). In contrast, if the local noise is symmetric, our scheme with $N$ chains, each of fixed length, will see a $1/\sqrt{N}$ improvement in measurement precision relative to standard metrology, and a $1/N$ improvement in the case of GHZ like entangled chains.

We also emphasise the robustness of our scheme to timing errors in the control operations.  Quantum metrology schemes typically use fast control operations that are highly susceptible to timing errors.  In particular, if one assumes Gaussian fluctuations in the timing of control pulses, this leads to a fidelity in unitary gate operations that decays exponentially in the variance of the fluctuations~\cite{Dur2007}.  In contrast, the unitary gates in our proposed scheme are adiabatic, and are completely insensitive to such timing errors. 

The main vulnerability of the scheme is to the presence of symmetry-breaking background fields.  We present simulations of the effect of perturbations in Fig.~\ref{fig:Fidelity}.  The simulations were performed on a spin chain of length $n=8$ where the initial state is the ground state of the Heisenberg interaction \emph{without perturbation} followed by measurement on the right edge in the $\ket{S^z=1}$ state so that the chain is initialized in the logical zero state $\ket{{\bf 0}}_{1,n}$. Next a local perturbation $O$ is turned on over the entire chain and the system evolved numerically using a Trotter time step $\delta t=0.01$ for a total time $T=10/J$ according to the time dependent Hamiltonian 
\[
  H(t)=\frac{t}{T} J (S^{x}_1)^2 +  (1-\frac{t}{T}) J \vec{S}_1\cdot \vec{S}_{2}+H_{2,n}+\gamma \sum_{j=1}^n O_j \,,
\]
resulting in the output state $\sigma$. Without perturbation, and in the adiabatic limit of zero excitation outside the ground space, the output state would be the input state experiencing a $\pi$ rotation about the $\hat{x}$ axis, i.e. the bit flipped state $\rho =\ket{{\bf 1}}_{2,n}\bra{{\bf 1}}_{2,n}$. The fidelity of the gate operation is then calculated as $F=(\tr[\sqrt{\sqrt{\rho}\sigma\sqrt{\rho}}])^2$.
  It is evident that the protocol is immune to symmetry respecting perturbations, but as expected if the perturbation breaks the $D_{\infty}$ symmetry, the fidelity degrades. However, we note that this degradation also occurs in a ``standard'' scheme with no protection.  We note that this might be ameliorated by actively incorporating dynamical decoupling schemes to remove such fields \cite{Viola:00,Bookatz:14}.

\section{Potential implementation with trapped Rydberg dressed atoms}

In \cite{BP:15}, it is shown how to simulate the Haldane model with spin-1 encoding in ground electronic states of Rydberg dressed atoms. The atoms are trapped one per well in an optical lattice or microtrap arrays with lattice spacings ranging from a half $\mu$m to a few $\mu$m. External lasers are applied to dress the ground electronic states of each atom with an excited Rydberg state such that each atom has an induced dipole moment. Multiple lasers can be used to dress with several Rydberg states with different quantum numbers in order to obtain tailored nearest neighbor interactions. Using atoms with an $F=1$ hyperfine ground state manifold such as $^{87}$Rb allows for engineering the spin-1 XYZ Hamiltonian and by tuning the ellipticity of the dressing lasers, the fully $SO(3)$ symmetric Heisenberg interaction that is required for our metrology scheme. 

The proof of principle simulation would begin by starting with all spins initialized in $\ket{F=1,m_F=0}$ then adiabatically turning on the dressing lasers to prepare the ground subspace of the Heisenberg chain on several atoms ($\sim 10$ would suffice \cite{RMBB2013}) and initializing a particular ground state by measuring boundary spins \cite{M2010,RMBB2013}. An edge mode could be manipulated by using a far off resonance, spin \emph{independent} optical tweezers trap that removes a single spin from the chain \cite{Beugnon:07, Kuhn:12}. An effective electric field local to the boundary spin can be realized using AC stark shifts from a laser with intensity $I$ detuned by $\Delta_L$ from the transition between a ground state $F=1$ manifold and excited state manifold. The strength of the effective electric field will be given by $J_f\sim \frac{Ic}{\Delta_L}$ where $c$ is the difference in the square of the Clebsch-Gordan coefficients of the ground to excited state magnetic sublevels, and the direction of the field is determined by the laser polarization which could be dynamically manipulated. Recently, single site addressing using this kind of AC stark shift was demonstrated in a 3D optical lattice for performing local quantum gates \cite{Wang:15}.

\section{Conclusions}

We have presented a robust scheme of metrology that uses the passive, error-preventing properties of SP topological phases.  A potential realization by Rydberg dressed atoms in optical lattices or microtrap arrays is promising, since other symmetry-breaking fields, like a magnetic field $\gamma\vec{S}$, have been routinely removed, to the $\mu$G level, in experiments using magnetic shielding \cite{Derevianko:11}. 

\begin{acknowledgements}
\textit{Acknowledgment.---}
SDB and GKB acknowledge support from the ARC via the Centre of Excellence in Engineered Quantum Systems (EQuS), project number CE110001013. AM acknowledges support from National Science Foundation grants PHY-1314955, PHY-1521016, and PHY-1620651.
\end{acknowledgements}

\end{document}